\documentclass[aps,prl,reprint,superscriptaddress,amsmath,amssymb,showkeys]{revtex4-2}
\usepackage{graphicx}
\usepackage{booktabs} 
\usepackage{framed} % Framing content
\usepackage{makecell}
\usepackage{bm}
\usepackage{physics}
\renewcommand{\vec}[1]{\boldsymbol{#1}} % vector
\newcommand{\phm}{\phi_\text{max}} % maximum-packing volume fraction
\begin{document}

%\preprint{}

\title{Magnetically induced deformation of isotropic magneto-active elastomers and its relation to the magnetorheological effect}


\author{Mehran Roghani}
\email[]{roghani@ipfdd.de}
\affiliation{Institute Theory of Polymers, Leibniz-Institut für Polymerforschung Dresden e. V.,\\ Hohe Straße 6, Dresden, 01069, Germany}
\affiliation{Faculty of Mechanical Science and Engineering, Dresden University of Technology, Dresden, 01062, Germany}
\author{Dirk Romeis}
\affiliation{Institute Theory of Polymers, Leibniz-Institut für Polymerforschung Dresden e. V.,\\ Hohe Straße 6, Dresden, 01069, Germany}

\author{Gašper Glavan}
\affiliation{East Bavarian Centre for Intelligent Materials (EBACIM), Ostbayerische Technische Hochschule (OTH) Regensburg}

\author{Inna A. Belyaeva}
\affiliation{East Bavarian Centre for Intelligent Materials (EBACIM), Ostbayerische Technische Hochschule (OTH) Regensburg}
\author{Mikhail Shamonin}
\affiliation{East Bavarian Centre for Intelligent Materials (EBACIM), Ostbayerische Technische Hochschule (OTH) Regensburg}

\author{Marina Saphiannikova}
\affiliation{Institute Theory of Polymers, Leibniz-Institut für Polymerforschung Dresden e. V.,\\ Hohe Straße 6, Dresden, 01069, Germany}
\affiliation{Faculty of Mechanical Science and Engineering, Dresden University of Technology, Dresden, 01062, Germany}

\date{\today}

\begin{abstract}
Can isotropic Magneto-Active Elastomers (MAEs) undergo giant magnetically induced deformations and exhibit huge magnetorheological effects simultaneously? In this experimental and theoretical study, we reveal how the macroscopic deformation of MAEs relates to the process of particle restructuring caused by application of a magnetic field. For this purpose, MAE cylinders with different aspect ratios and particle loadings are studied in uniform magnetic fields. The axial deformations of the cylinders are acquired using an optical camera. A unified mean-field model proposed in previous studies is adapted to describe the transition of initially isotropic cylinders into transversely isotropic ones. This mechanical transition is caused by the rearrangement of particles into dense columnar structures aligned with the field and is believed to result in a huge magnetorheological effect. Our model however predicts less than three-fold increase in elastic moduli when evaluated along the field direction. This prediction is based on a careful examination of the shear moduli of studied MAEs and the columnar structures. A weak magnetorheological effect explains significant axial deformations measured in the field direction. A strong magnetorheological effect would hinder axial deformations due to an increase in modulus by several orders of magnitude. Not only are the moduli and macroscopic deformations influenced by microstructure evolution, but so is the magnetization of particles, which increases as they rearrange into dense columns. With this study, we show that the unified mean-field model provides quantitative access to hidden material properties such as magnetization and stiffness in MAE samples with different shapes and evolving microstructures.
\end{abstract}


\maketitle


\section{Introduction}

The advancement of composite materials that offer the ability of tailoring their properties through material design and external stimuli has set a new trend in engineering fields \cite{jain2023smart}. Among these, Magneto-Active Elastomers (MAEs) have gained much attention by providing a wide range of customization options. Their microstructure can be designed into many different forms during the fabrication \cite{ginder1999magnetorheological, bastola2018development,bastola2017novel, zhang2020magneto, qi20203d, bastola2020dot, bastola2020line, haussener2021particle} and can also be altered remotely after fabrication using external magnetic fields \cite{borin2023use, gundermann2014investigation, SchümannOdenbach2023}. The remotely induced microstructure alterations arise from the magnetic interactions between the dispersed particles within the elastomeric matrix and can make the mechanical, magnetic and electrical properties of MAEs programmable \cite{snarskii2019theoretical, savelev2024enhancement, kostrov2023magnetoactive}. The most practically exploited aspect of MAEs is their ability to undergo magnetically induced deformations and changes in elastic moduli (magnetorheological effect) when exposed to magnetic fields. This makes MAEs exceptionally promising for innovative applications like soft robotics \cite{reiche2023multipole, becker2022magnetoactive, alapan2020reprogrammable}, medical devices \cite{zhang2019magnetic, calvo2019biocompatible} and responsive surfaces \cite{straus2024surface, kriegl2024tunable}.

The particles used in the fabrication of MAEs can be hard or soft magnetic materials, according to their magnetization behavior, which can either exhibit pronounced magnetic hysteresis, e.g. NdFeB, or not, e.g. carbonyl iron \cite{moreno2022effects}. The polymer network of the matrix in MAEs helps to retain the structural integrity of the sample. The magnetic particles maintain their fixed positions in an MAE after curing, unlike magnetorheological fluids where the shape is not fixed and particle sedimentation occurs. However, MAEs can show significant changes in mutual arrangement of particles when exposed to magnetic fields. This phenomenon of microstructure evolution is most pronounced in MAEs based on soft elastomeric matrices. Many studies have visualised the evolution of the microstructure using X-ray micro-computed tomography and observed how the randomly distributed particles form columnar structures aligned along the direction of the magnetic field lines  \cite{gundermann2014investigation, chen2022situ, urano2023extremely}. One of the effects of this phenomenon is a noticeable increase in the Young's modulus of MAEs along the magnetic field compared to an almost unchanged modulus in the perpendicular direction \cite{jaafar2021review,bodelot2018experimental}. The other effect is an increase in the shear modulus of MAEs, when shear deformation is applied perpendicular to the magnetic field. These magnetorheological effects reflect the appearance of mechanical anisotropy (transverse isotropy) introduced by a magnetic field in originally isotropic samples \cite{chougale2021transverse,spaggiari2021magnetorheological}. Several experimental studies have been conducted in the past years to explore the degree of the magnetorheological effects and have shown an increase in moduli ranging from tens of percents to several orders of magnitude \cite{bastola2020recent, boczkowska2012mechanical, moreno2021new, kostrov2024influence, yao2018magnet, stoll2014evaluation, nanpo2016magnetic, sorokin2014experimental}. 
The highest magnetorheological effect over several orders of magnitude was observed for ultra-soft MAEs, where the effective shear modulus was of the order of 1 kPa and lower, see, e.g., Table 1 in \cite{stoll2014evaluation}.

Many theoretical works have investigated the magnetic and mechanical behaviour of MAEs \cite{nadzharyan2022theoretical, saber2023modeling, kalina2024neural, kalina2023multiscale, moreno2023influence, fischer2019magnetostriction, goh2023density, stolbov2024striction, zubarev2015effect, dobroserdova2023switching}, with some aiming to comprehend the origin of the huge magnetorheological effect \cite{chougale2022magneto, snarskii2021effect, tsai2016tailoring, ivaneyko2018dynamic, garcia2021influence, fischer2024magnetic}. There are several studies in which experimental and theoretical approaches are combined, predominantly focusing on stiff 
(initial shear modulus $\sim 10^5 - 10^6$ Pa) samples and employing phenomenological models \cite{ardehali2021characterization, bellelli2019magneto, shen2004experimental, koo2010dynamic}.Taking into account magnetic interactions between the particles, even when rearranged into columnar structures, leads to a weak increase in both the elastic and shear moduli \cite{ivaneyko2018dynamic, chougale2021transverse}. In recent years, a hypothesis was put forward that the formation of clusters with increased particle density and thus increased elastic modulus along the magnetic field lines is responsible for the huge effects \cite{snarskii2021effect}.

To prove this hypothesis, in a previous work we introduced a unified mean-field model that considers the effects of microstructure evolution on the mechanical behavior of MAEs \cite{roghani2023effect}. There, we elaborated that a field-induced increase in the shear modulus over several orders of magnitude observed in some experimental setups is only possible if the sample undergoes significant microstructure evolution. We also showed that this would in turn hinder the deformation of the sample, simply because the stiffness increases dramatically. Interestingly, many experimental studies on soft (effective shear modulus in the absence of a magnetic field $\approx 30 - 50$ kPa) MAEs demonstrate significant elongations of the samples along the magnetic field direction \cite{silva2022giant, saveliev2020giant, stepanov2013magnetodeformational}.

Recently, Glavan et al. \cite{glavan2023experimental} measured quasi-static deformations of isotropic soft MAE cylinders and found that axial deformation of up to 35\% can be achieved with moderate magnetic fields. They also complemented that study by examining the transient deformations of the cylindrical MAEs \cite{glavan2024transient}. These measurements also suggest that the stiffness of soft, unconstrained MAEs is unlikely to increase over many orders of magnitude in the presence of technically feasible magnetic fields.
Indeed, direct measurements of the shear modulus on disc-shaped MAE films constrained along magnetic field demonstrated that the maximum magnetorheological effect was about one order of magnitude for the strain amplitude of 0.01\% and it drastically decreased with increasing strain amplitude of shear oscillations (see, Fig. 4 in \cite{glavan2023experimental}).  

In the present work, we build on our unified mean-field model \cite{roghani2023effect} and adapt it in order to be compatible with experimental measurements on magnetically induced axial deformations of MAE cylinders under uniform field loadings. This gives us the opportunity to investigate the interrelated changes in deformation and stiffness of MAEs in direct comparison with corresponding experimental findings. To this end, eight cylindrical MAE samples are manufactured with various aspect ratios and volume fractions of carbonyl iron powder (CIP). By fitting the field-induced deformations of MAE samples to our unified mean-field model, we gain insight on what range of stiffness changes is possible under magnetic loading. In addition, we include predictions on the effects of microstructure evolution on the magnetization behaviour of CIP particles within the sample.

\section{Unified mean-field model}

\begin{figure}[ht]
	\centering
	\includegraphics[scale=.25]{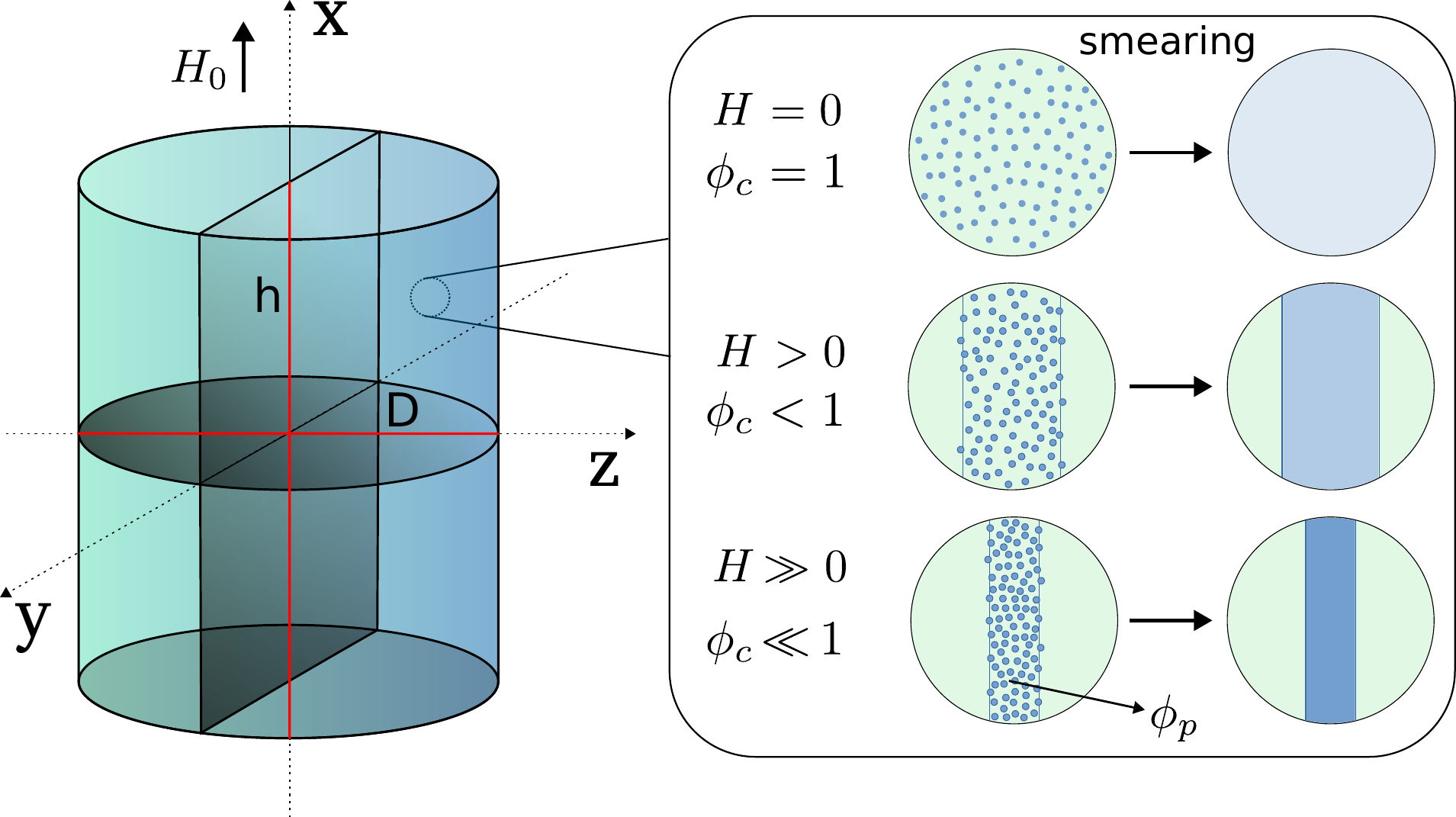}
	\caption{Left: Schematics of an MAE cylinder and its axes, h and D. Right: Schematics of how the microstructure evolution under the local magnetic field $\vec{H}$ is captured in our model. The randomly distributed particles attract each other and form cylindrical particle-rich regions (columns). As the magnitude $H$ of the magnetic field increases, particles move closer to each other and the volumetric portion of the columns $\phi_c$ with respect to the sample volume decreases while the local volume fraction $\phi_p$ increases. Smearing shows how the particle and matrix properties (magnetic and mechanic) are averaged over these regions to simplify the calculations.}
	\label{gesch}
\end{figure}

This section addresses the details of our model for magneto-mechanical behavior of cylindrical MAE samples. We denote this model in the following as unified mean-field approach, because it simultaneously accounts for the elastic (effective medium approximation) and the magnetic (dipolar mean-field) properties in a self-consistent field approach. Thereby, our model allows to unify the most relevant effects on macroscale, i.e., sample shape and deformation, and on microscale, i.e., particle arrangement and restructuring. In the following, the cylinder is considered to be in a uniform magnetic field, with its symmetry axis always parallel to the direction of external magnetic field $\vec{H}_0$ (depicted in the left part of Fig. \ref{gesch}).

We define the free energy density of this cylinder, $\Psi_{MAE}$, as consisting of three components \cite{roghani2023effect}:
\begin{equation}\label{PMAE}
	\Psi_{MAE}(\vec{F},\vec{H})=\Psi_{mech}(\vec{F},\vec{H})+\Psi_{mag}(\vec{F},\vec{H})+\Psi_{mic}(\vec{H}).
\end{equation}
The mechanical energy density $\Psi_{mech}$ accounts for macroscopic deformations and is a function of the deformation gradient $\vec{F}$ and the magnetic field $\vec{H}$. The dependence of mechanical energy on the magnetic field is due to the macroscopic implications of the microstructure evolution, explained in detail in the next section. The magnetic energy density $\Psi_{mag}(\vec{F},\vec{H})$ accounts for dipolar interactions between the magnetized particles. The dependence of magnetic energy on the deformation gradient is related to the changes in the demagnetizing factor of the cylinders caused by the deformation. $\Psi_{mic}(\vec{H})$ represents the microscopic mechanical energy density, which addresses the energy penalty required for microstructure alterations, or in other words, the energy associated with the local stretching of the elastomeric matrix by the inclusions. In the following, we provide a detailed description of each of these three components.

\subsection{Mechanical energy}

As explained in the introduction, the application of an external magnetic field causes a transition of the MAE material from its initial isotropic state to an anisotropic one. This transition leads to a large increase in the elastic moduli along the direction of the magnetic field in comparison to the moduli perpendicular to that direction (this type of anisotropy is also called transverse isotropy). Anisotropy is weak at low magnetic fields and becomes stronger as the magnetic field increases.

To describe the mechanical energy considering this transitional behavior from isotropic to anisotropic, we implement a modified Neo-Hookean model:
\begin{equation}\label{psimec}
	\Psi_{mech}=\frac{G_{iso}}{2}\left[(\zeta_1+1)(I_1-3)+\zeta_2(I_4-1)^2\right],
\end{equation}
where the effective shear modulus of the isotropic composite is defined by $G_{iso}$. Dimensionless anisotropy parameters $\zeta_1$ and $\zeta_2$ affect the mechanical properties in the two directions mentioned before: along the magnetic field and perpendicular to it. We use the first invariant $I_1=\Sigma \lambda_i^2$ and the pseudo-invariant $I_4=\lambda_1^2$ derived in our previous work \cite{roghani2023effect} for the case of uniaxial deformation. Both invariants are defined in terms of the stretch ratios $\lambda_i$, where $i$ takes the values 1, 2 and 3 corresponding respectively to the principal axes $x, y$ and $z$.

In the macroscopic continuum approaches, field-induced anisotropic behavior of initially isotropic MAEs has traditionally been modeled using six invariants \cite{kankanala2004finitely, kalina2020macroscopic, danas2024electro}, where the invariant $I_4$ is purely magnetic, while the invariants $I_5$ and $I_6$ capture the magneto-mechanical coupling. Compared to these macroscopic approaches, our unified mean-field model explicitly considers dipolar interactions between the magnetized particles which is described in the next section. This allows to bridge the microscopic and macroscopic scales \cite{metsch2021magneto,romeis2021cascading} by taking into account both the local particle distribution and the sample shape. The advantage of our approach is that the magnetic energy density is described in the framework of dipole approximation and hence there is no need to introduce phenomenological invariants into its expression. 

Introducing $\gamma_0=h/D$ as the initial aspect ratio of the cylinder, the aspect ratio of the deformed but incompressible sample can be written in terms of the axial stretch ratio $\lambda_1$:
\begin{equation}\label{aspcr2}
	\gamma=\gamma_0 \lambda_1^{\frac{3}{2}}.
\end{equation}

The anisotropy parameters $\zeta_1$ and $\zeta_2$ are evaluated for each stage of the microstructure evolution according to the particle volume fraction inside the columnar structures. In the absence of magnetic field, the particles are randomly distributed in the sample (assuming the sample is fabricated without a curing magnetic field). Therefore, isotropic mechanical properties in the absence of magnetic field are achieved by initially setting both anisotropy parameters to zero. In the presence of a magnetic field, the magnetized particles rearrange themselves into columnar structures aligned with the magnetic field lines \cite{gundermann2014investigation}. This behavior is formalized here by assuming the formation of two regions on microscale, i.e. particle-rich columns and the unfilled matrix that surrounds these columns (right part of Fig. \ref{gesch}). The volume fraction of the columnar regions in a sample is denoted by $\phi_c$ and the local particle volume fraction inside these columns is expressed as $\phi_p$. These two volume fractions are linked to the total volume fraction of particles inside the sample $\phi$ with the equation below:
\begin{equation}\label{fipi}
	\phi_c=\frac{\phi}{\phi_p}.
\end{equation}

In the absence of magnetic field, $\phi_c=1$ and therefore the isotropic sample can be represented as one column with $\phi_p=\phi$. By increasing the magnetic field, $\phi_p$ starts to increase, and this leads to the transition of the material from isotropic to anisotropic state due to the formation of columnar structures aligned with the field. In this way, $\phi_p$ acts as a mean-field order parameter in our model. It should be noted that we consider the columns to have isotropic properties \cite{chougale2022magneto}. The presence of aligned columns is more favorable than a random distribution in terms of the magnetic energy. This phenomenon is explained in more detail in the section dedicated to the magnetic energy. The transition towards more and more dense columns impacts the mechanical energy density by inducing a change in the anisotropy parameters values so that the elastic modulus of the material along the magnetic field direction is largely increased compared to the modulus perpendicular to the field.

Further, we need an effective medium theory that is capable of determining the isotropic shear modulus of our composite and also that of the columns. Recently, Lef{\`e}vre and Lopez-Pamies \cite{lefevre2022effective} presented a new formula describing the effective shear modulus of composites with spherical inclusions:
\begin{widetext}
\begin{equation}\label{EMTLP}
	G_e=\frac{G_m}{\left[\left(1+\alpha\left(\beta+\left(\frac{\phi}{\phm}\right)^2\right)\left(\frac{\phi}{\phm}\right)^2\right)\left(1-\frac{\phi}{\phm}\right)\right]^{\frac{5\phm}{2}}},
\end{equation}
\end{widetext}
where $\alpha$ and $\beta$ are two fitting parameters and $\phm$ represents tight packing volume fraction of the particles. The shear modulus of the elastomeric matrix is $G_m$ and $\phi$ is the volume fraction of inclusions. Equation \ref{EMTLP} is a modification of the well-known relation, that Krieger and Dougherty obtained in the style of mean-field approach by extending the dilute limit results for shear viscosity to high volume fractions \cite{Krieger1959}. As we showed previously \cite{Ivaneyko2022}, the same Krieger-Dougherty relation can be applied to the composites based on a Neo-Hookean matrix:
\begin{equation}\label{KriegDoug}
	G_e=\frac{G_m}{\left[1-\frac{\phi}{\phm}\right]^{[\mu]\phm}}.
\end{equation}
Here, $[\mu]$ is the intrinsic modulus defined by the shape of a particle. We note that Eq. \ref{EMTLP} with $\alpha=0$ reduces to the Krieger-Dougherty relation (Eq. \ref{KriegDoug}), when $[\mu]=5/2$ is substituted for rigid spherical particles. The advantage of the new formula (\ref{EMTLP}) proposed by \citet{lefevre2022effective} is that the presence of two additional parameters $\alpha$ and $\beta$ provides greater flexibility in describing the reinforcement of an elastomeric matrix in real composite materials.

This formula is then fitted to the shear (storage) moduli of MAE samples with volume fractions $0$, $0.22$, $0.27$ and $0.33$ reported in a previous work \cite{glavan2023experimental} to find the parameters $\alpha$ and $\beta$. Since the volume fraction of tight packing for monodisperse inclusions is estimated to be $\phm=0.64$, a value of $\phm=0.7$ seems to be reasonable for our samples filled with polydisperse inclusions. The best fit is found using the parameters listed in Table \ref{EMTt}.

\begin{table}[b]
\caption{\label{EMTt}%
Parameters used to fit the shear moduli of MAE samples to Eq. \ref{EMTLP} and their values.
}
\begin{ruledtabular}
\begin{tabular}{cc}
\textrm{Parameter}&
\textrm{Value}\\
\colrule
$\phm$&0.7\\
$\alpha$&1\\
$\beta$&-2\\
\end{tabular}
\end{ruledtabular}
\end{table}

\begin{figure}[ht]
	\centering
	\includegraphics[scale=.8]{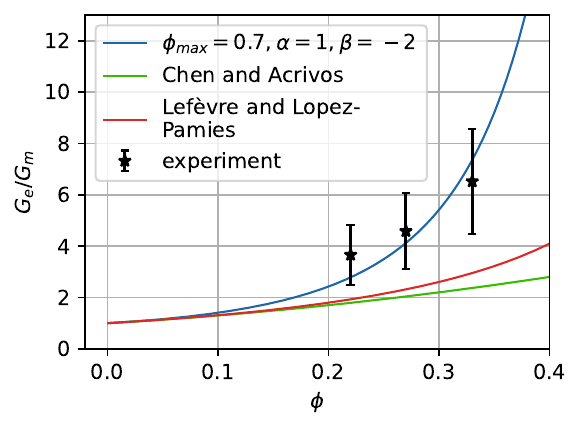}
	\caption{Normalized shear modulus versus the volume fraction of particles. Blue line: Eq. \ref{EMTLP} fitted to the shear moduli of MAE samples with different volume fractions. Red line: the fit obtained by \citet{lefevre2022effective} with 3D finite element calculations for monodisperse inclusions with $\alpha=0.635, \beta=0.027, \phm=0.64$. Green line: the dilute limit results by \citet{hsiao1978effective}. Black stars with error bars show the experimentally measured shear moduli of samples with different volume fractions and their relative uncertainty \cite{glavan2023experimental}.}
	\label{EMT}
\end{figure}

Fig. \ref{EMT} shows the fit acquired from Eq. \ref{EMTLP} compared to the fit obtained to the 3D finite element calculations for monodisperse inclusions in \cite{lefevre2022effective} ($\alpha=0.635, \beta=0.027, \phm=0.64$ are the parameters used in \cite{lefevre2022effective}) along with the dilute limit results of \citet{hsiao1978effective}.

As it can be seen from Fig. \ref{EMT}, the results of \cite{hsiao1978effective} and \cite{lefevre2022effective} have the same initial slope. This initial slope (around $\phi=0$) can be obtained by a Taylor expansion of Eq. \ref{EMTLP}, which results in the following expression for the amplification factor: 
\begin{equation}\label{EMTLP_expanded}
X=G_e/G_m=1+5/2\phi+b\phi^2+O(\phi^3).
\end{equation}
From the fit of shear moduli for MAE samples, we obtain the parameter $b=(10+25\phi_{max}-20\alpha\beta)/8\phi_{max}$ which takes a value around $12$, noticeably higher than the value $5.0$ predicted by \citet{hsiao1978effective}. The reason behind this difference could be that the results of \cite{hsiao1978effective} and \cite{lefevre2022effective} only take into account the effect of hydrodynamic reinforcement, neglecting the effects of interactions between inclusions or depletion interactions with the polymer matrix, which lead to the agglomeration of particles \cite{vilgis2009reinforcement}. Also, a so-called localized layer is usually formed in the vicinity of the particle surface. Below the glass transition, this localized layer is considerably stiffer compared to the rest of elastomeric matrix which is found further away from the particles. In the absence of agglomeration, the amount of localized layer can be estimated using the hard core/soft shell model \cite{vilgis2009reinforcement} or the superposition approach \cite{ivaneiko2017superposition}, and can be used to modify an expression for the amplification factor. It is worth mentioning that a value of $b=14.1$ (Guth-Gold model \cite{zapp1951elastic}) is frequently used in the industry \cite{domurath2017concept}. Recently, Nadzharyan and Kramarenko \cite{nadzharyan2023effects} found an amplification factor close to Guth-Gold model for MAEs using a single inclusion modeling approach. Since multiple intertwined effects can influence the amplification factor, we prefer to use phenomenological Eq. \ref{EMTLP} which reasonably fits the shear moduli of our MAE samples. 

Next, we determine the elastic moduli of the composite along the direction of external magnetic field, $E_L$, and perpendicular to it, $E_T$. The rules of mixtures are used to obtain these two moduli as follows:
\begin{equation}\label{eq7}
	E_L=(1-\phi_c)E_m+\phi_cE_c,
\end{equation}
\begin{equation}\label{eq8}
	\frac{1}{E_T}=\frac{1-\phi_c}{E_m}+\frac{\phi_c}{E_c},
\end{equation}
where $E_m=3G_m$ and $E_c=3G_c$, according to the known relation $E=3G$ for isotropic incompressible materials. The modulus of the isotropic columns $G_c=G_e(\phi_p)$ is obtained from Eq. \ref{EMTLP} by replacing $\phi$ with the local volume fraction $\phi_p$. 

We can also calculate $E_L$ and $E_T$ by taking the second derivative of the mechanical energy density (Eq. \ref{psimec}):
\begin{equation}\label{D2mch1}
	\overline{E_L}=\eval{\frac{\partial^2 \Psi_{mech}}{\partial \lambda_1^2}}_{\lambda_1(H_0)},
\end{equation}
\begin{equation}\label{D2mch2}
	\overline{E_T}=\eval{\frac{\partial^2 \Psi_{mech}}{\partial \lambda_2^2}}_{\lambda_1(H_0)}.
\end{equation}
These derivatives become functions of the anisotropy parameters, $f(\zeta_1, \zeta_2)$. Subsequently, by equating the values of $E_L$ and $E_T$ obtained from the rules of mixture to the ones obtained by differentiation of the mechanical energy:
\begin{equation}\label{D2mch1eq}
	\overline{E_L}(\zeta_1, \zeta_2)=E_L,\quad \overline{E_T}(\zeta_1, \zeta_2)=E_T,
\end{equation}
the values of anisotropy parameters at each stage of microstructure evolution are calculated.

All of these steps render the elastic energy density into a function of two free variables: the axial stretch ratio $\lambda_1$ and the local volume fraction $\phi_p$. The shear modulus of the elastomeric matrix $G_m$ and the volume fraction of particles within the sample $\phi$ are kept constant.

\subsection{Magnetic energy}

In the following we assume that the MAE sample is homogeneously magnetized, i.e. each particle adopts the same magnetization, directed along the external magnetic field $\vec{H_0}$. The magnetic energy density then takes the form \cite{roghani2023effect}:
\begin{equation}
	\Psi_{mag}=\mu_0\phi\left(-\int_{0}^{H}M\mathrm{d} H +\frac{1}{2}M(H-H_0)\right),
\end{equation}
where $\mu_0$ indicates the permeability of vacuum, and $H$ and $M$ denote the local magnetic field and magnetization both directed along $\vec{H_0}$. In our previous paper \cite{roghani2023effect}, the magnetic homogeneity assumption was guaranteed by considering an ellipsoidal shape of the sample. Here, we describe a cylindrical sample form where field inhomogeneity does emerge, especially in the regions around the edges at the top and the bottom of the cylinder \cite{romeis2021cascading}. In this work, we aim to formulate a unified model that incorporates effects in a leading-order approximation by neglecting field inhomogeneities. The model is capable of describing the most relevant changes in behavior when a magnetic field is applied and allow for a quantitative comparison with experimental data.

The magnetization in the particles is related to the local magnetic field through a magnetization function specific to the material from which the particles are made of. Here, we consider carbonyl iron powder and assume that the magnetization of individual particles follows a Langevin function that captures its saturating nature \cite{romeis2023effective}:
\begin{equation}\label{MagFlan}
	M(H)=M_{\infty}\left[\coth{\frac{3\chi}{M_\infty} H}-\frac{1}{\frac{3\chi}{M_\infty} H}\right],
\end{equation}
where $\chi$ and $M_\infty$ denote the initial magnetic susceptibility and saturation magnetization of the particles, respectively. The saturation magnetization of the CIP particles used here was determined in an earlier work \cite{glavan2021magnetoelectric} to be $M_\infty=1659$ $\mathrm{kA m^{-1}}$. Also, based on the work of Ivaneyko et al. \cite{ivaneyko2011magneto}, we consider the initial susceptibility of the particles to take a relatively high value of $\chi=130$. According to theoretical modeling done by Romeis and Saphiannikova \cite{romeis2023effective}, taking a higher value for particle susceptibility would not noticeably affect the susceptibility of the composite. 

Inside the sample, the local magnetic field is directed along the external field $\vec{H_0}$, and thus along the symmetry axis of the cylinder. It takes the form:
\begin{equation}
	H=H_0-N M,
\end{equation}
where $N$ is the generalized demagnetizing factor proposed previously by Romeis et al. \cite{romeis2020magnetic}. This factor is concisely expressed by two parameters, which account for the shape of a sample, $f_{macro}$, and its microstructure, i.e.\ the local arrangement of particles, $f_{micro}$:
\begin{equation}\label{Hd}
	N=1/3-\phi f_{macro}-f_{micro}.
\end{equation}
where $f_{macro}=1/3-N_{\parallel}$ is a function of the demagnetizing factor of a cylinder magnetized homogeneously along its symmetry axis, $N_{\parallel}$. \citet{sato1989simple} presented a generic analytical expression for this demagnetizing factor reading:
\begin{equation}\label{demag}
	N_{\parallel}=1/(4\gamma/\sqrt{\pi}+1),
\end{equation} 
where $\gamma$ is the aspect ratio of the cylinder. \citet{sato1989simple}
showed that this simple expression predicts a demagnetizing factor for a homogeneously magnetized cylinder that is different from the exact value by less than 4.5\%. As shown in Eq. \ref{aspcr2}, $\gamma$ is related to the deformation of the cylinder because it is a function of the axial stretch ratio. In this way, the magnetic energy density contributes to the deformation of the sample.

The microstructure factor $f_{micro}$ is estimated based on the particle restructuring mechanism explained in the previous chapter (right part of Fig. \ref{gesch}) and mean-field approximation \cite{romeis2023effective} via the analytical form:
\begin{equation}\label{fmicro}
	f_{micro}=\frac{\phi_p-\phi}{3},
\end{equation} 

For isotropic samples $f_{micro}=0$, and it increases as the particles arrange more and more into anisotropic columnar-like structures. This indicates that the magnetic energy decreases as anisotropy increases and thus we can deduce that very dense columns are always favorable in terms of the magnetic energy.

These steps render the magnetic energy as a function of the axial stretch ratio $\lambda_1$ of the cylinder through $f_{macro}$ and the local volume fraction $\phi_p$ in columnar structures through $f_{micro}$. The external magnetic field parameter ($H_0$) is prescribed to actuate the modeled sample.

\subsection{Micro-mechanical energy}

The elastic energy $\Psi_{mech}$ presented in section 2.1., only takes macroscale deformations into account. As we explained previously, particle restructuring into columns is magnetically favorable. This results in very dense columns forming immediately at any magnetic field strength if there is no mechanical energy penalty for microstructure evolution. To overcome this issue, we proposed a micro-mechanical energy function for both initially isotropic and anisotropic magneto-active elastomers in our previous work \cite{roghani2023effect}. A reduced form of that energy function for initially isotropic samples takes the Hookean form:
\begin{equation}\label{microel}
	\Psi_{mic}= \frac{k_m}{2} G_{iso} f_{micro}^2,
\end{equation}
where the microscale deformations caused by the particle rearrangements are described on the basis of the previously defined variable $f_{micro}$ (Eq. \ref{fmicro}). This energy relies on the initial shear modulus $G_{iso}$, since the stiffness of the medium the particles are surrounded by affects the elastic energy needed for them to rearrange. Additionally, an adjustable micro-mechanical parameter, $k_m$, is introduced in Eq. \ref{microel} to control the degree of microstructure evolution. Setting $k_m$ to zero allows instant forming of dense columns under any magnetic fields. On the other hand, setting this parameter to a very high value will freeze the initial microstructure for any magnetic loading.

\section{Experiment}

Our MAE samples are based on a polydimethylsiloxane (PDMS) matrix according to a standard procedure outlined previously \cite{stepanov2014magnetoactive,glavan2023experimental} and CIP particles (type SQ, BASF SE Carbonyl Iron Powder and Metal Systems, Ludwigshafen, Germany) with a mean size of 3.9–5.0 $\mu$m. The matrix had a chemical composition corresponding to a stoichiometry ratio $r\approx1$, representing the ratio of molar concentrations of hydride and vinyl reactive groups. Ideally, only elastically active chains should be present in the polymer network.

Various materials were used to produce the soft elastomer matrix, including base polymer VS 100000, chain extender modifier 715, reactive diluent polymer MV 2000, crosslinker 210, Pt-catalyst 510, and inhibitor DVS, all provided by Evonik Operations GmbH, Geesthacht, Germany. A silicone oil (WACKER\textsuperscript\textregistered AK10) acted as a plasticizer. The cylinders were made with different CIP volume fraction $\phi$ and aspect ratios $\gamma_0=h/D$. The diameter $D$ was kept constant at 15 mm, while the height $h$ was varied between 3 and 18 mm with steps of 3 mm. The moulds used for the fabrication of the cylinders were 3D printed with heat resistant acrylonitrile butadiene styrene (ABS). Part of the MAE material was prepared for rheological characterization. Those rheological measurements were conducted using a commercial rheometer (Anton Paar, model Physica MCR 301) at a fixed angular frequency $10$ s$^{-1}$ and shear oscillation amplitude $0.01$\% \cite{glavan2023experimental}.

The cylinders were placed between the poles of an electromagnet (EM2 model, MAGMESS Magnetmesstechnik Jürgen Ballanyi e.K., Bochum, Germany), which generated a vertical magnetic field. The magnetic field was highly uniform over the 34 mm gap \cite{glavan2021magnetoelectric}. The diameter of magnetic poles was 92 mm. The electromagnet was powered by a bi-polar power supply (FAST-PS 1k5, CAENels s.r.l., Basovizza, Italy). The dimensional changes of cylinders were recorded using a CMOS camera (Alvium 1800 U-319 m, Allied Vision Technologies GmbH, Stadtroda, Germany) with a suitable lens (Edmund Optic Double Gauss Focusable, 25 mm C-mount F4.0 1.300, Barrington, NJ, USA). To achieve the highest possible contrast, samples were backlight illuminated with a light emitting diode (LED, Illuminant G4 Pen, Conrad Electronics, Hirschau, Germany) and a diffuser (Perspex diffuse, 2.5 mm, 3A Composites GmbH, Sins, Switzerland). The experiment was automated using LabVIEW software (version 2018 National Instruments, Austin, TX, USA). The vertical position of the camera was adjusted for each cylinder in such a way that the viewing angle on the upper edge of a cylinder was kept constant regardless of the sample’s aspect ratio \cite{glavan2023experimental}.

\begin{figure}[ht]
	\centering
	\includegraphics[scale=.3]{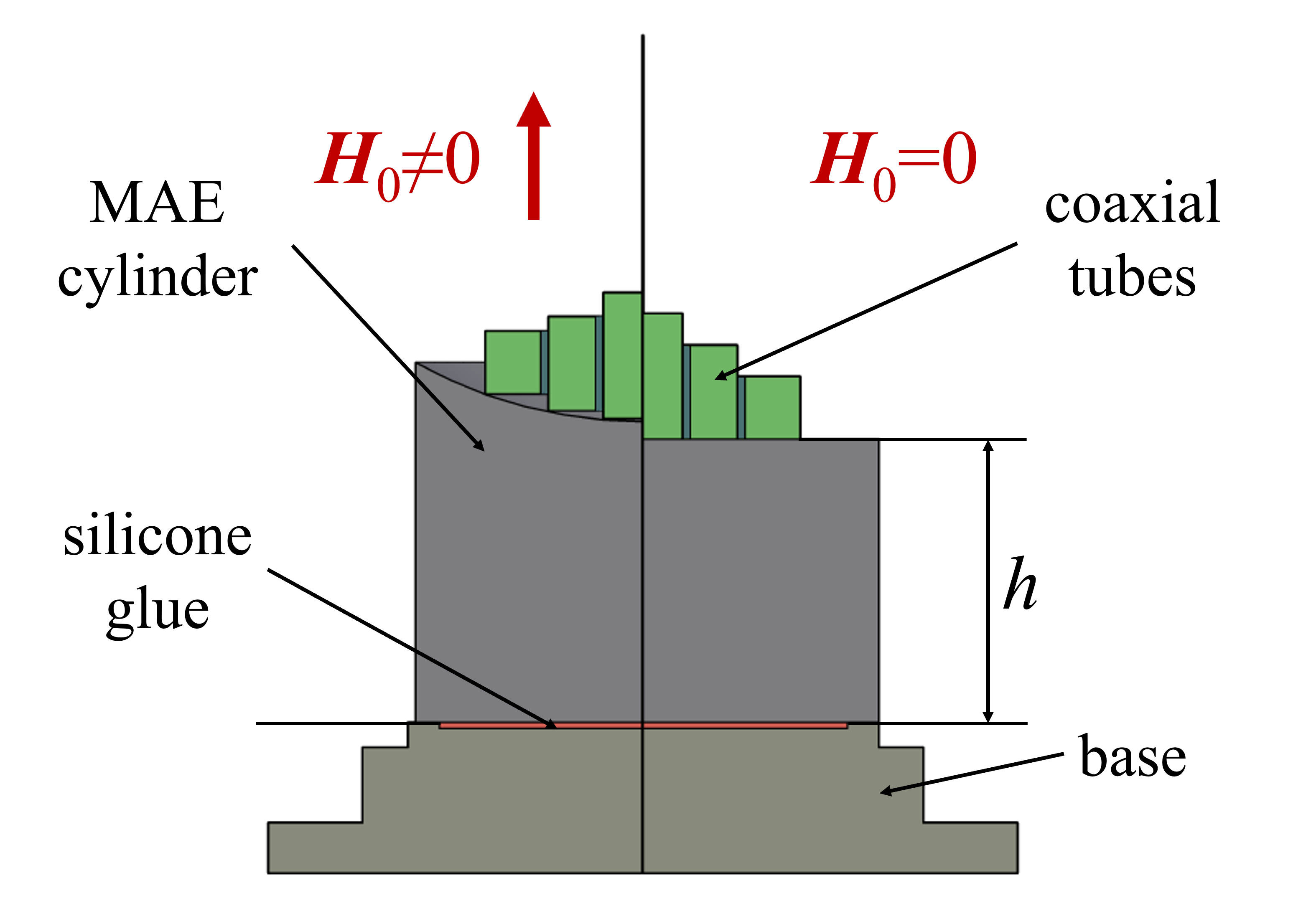}
	\caption{Schematic of the cross-section of MAE cylinder with co-axial tubes in magnetic field $H_0$ (left) and in zero field (right).}
	\label{schem}
\end{figure}

Upon applying the external magnetic field, a concave dent is formed on the top surface of cylinder. In order to characterise the dent, three co-axial thin-walled concentric tubes of different diameter and different height were positioned on the upper cylinder surface (Fig. \ref{schem}). With the deformation of the upper surface, the co-axial tubes were pushed upwards, making the dents shape deducible from the displacement of the tube’s top edges \cite{glavan2023experimental}.

The captured images were analysed using a Python script with OpenCV library. The fixed (bottom) edge of the cylinder was set manually. The other edges were followed using the Cunny edge detection algorithm \cite{glavan2023experimental}.

From the relative change in heights of the concentric tubes, the displacement of the top surface is measured in 5 points. These values and the displacement of the edge of the cylinder are then used to calculate the volume of the convex part on the top. Dividing this volume by the cross sectional area of the deformed cylinder results in the average deformation value. This average would be the displacement of the cylinder if the top surface stayed flat during the application of the magnetic field. We believe that using this average deformation provides a straightforward and effective measure of deformation.

The electrical current $I$ in the electromagnetic coils was adjusted between 0 and 10 A to vary the magnetic field strength corresponding to external magnetic field $H_0$ between 0 and 485 kA/m. The electrical current was changed in steps of 0.5 A ($\approx$25 kA/m) in 20 s intervals, which was long enough for cylinders deformation to reach a steady state ($\tau\approx 0.5$ s). Measurements were conducted at room temperature \cite{glavan2023experimental}.

\section{Results and discussion}

To predict the magnetically induced deformation of cylindrical MAEs, we minimize the free energy density (\ref{PMAE}) with respect to the free variables: the axial stretch ratio of the cylinder ($\lambda_1$) and the particle volume fraction inside the columns ($\phi_p$). This is done at each value of the applied magnetic field using empirical parameters shown in Table \ref{tbl1}. 

\begin{table*}[t]%The best place to locate the table environment is directly after its first reference in text
\caption{\label{tbl1}%
Empirical model parameters and their values.
}
\begin{ruledtabular}
\begin{tabular}{cccc}
\textrm{Parameter}&
\textrm{Description}&
\textrm{Value}&
\textrm{Unit}\\
\colrule
$G_m$&\makecell{matrix shear modulus}&7.7 & kPa\\
$M_\infty$&\makecell{saturation magnetization of the CIP particles}&1659&kA/m\\
$\chi$&\makecell{magnetic susceptibility of the CIP particles}&130&-\\
\end{tabular}
\end{ruledtabular}
\end{table*}

The shear modulus of the matrix $G_m$ as well as the shear moduli of MAEs with different volume fractions were measured dynamically (at $\omega=10 \quad s^{-1}$) with a rheometer, as described in \cite{glavan2023experimental}. In this paper we refer to the storage modulus simply as shear modulus and do not consider the loss modulus which is significantly lower. Because of the dynamic nature of the testing procedure, we anticipate that these shear moduli are higher than the moduli of the cylindrical samples deformed quasi-statically in this study. It is well known that the dynamic moduli of elastomers increase with the testing frequency \cite{saphiannikova2014multiscale}. During the deformation measurement, the magnetic field is applied in step-wise increments each lasting 20 seconds, which results in a quasi-static behavior. Also, as it can be seen in Fig. \ref{EMT}, the experimental measurements of shear modulus have a large margin of error. This further complicates the assignment of the value of $G_m$ in our model. To compensate for these impacts, we introduce a correction coefficient, $k_e$, for the shear modulus $G_m$. This modification does not affect the parameters $\alpha$, $\beta$ and $\phm$ used to fit the effective medium theory in Fig. \ref{EMT}.

The shear correction coefficient $k_e$ and the micro-mechanical parameter $k_m$ are tuned to find a good fit to the experiment. The values of these fitted parameters are presented in Table \ref{tbl2}.

\begin{table}[b]%The best place to locate the table environment is directly after its first reference in text
\caption{\label{tbl2}%
Fitted model parameters and their values for the fit to experiment.
}
\begin{ruledtabular}
\begin{tabular}{ccc}
\textrm{Parameter}&
\textrm{Description}&
\textrm{Value}\\
\colrule
$k_m$&\makecell{micro-mechanical parameter}&550\\
$k_e$&\makecell{shear correction coefficient}&0.8\\
\end{tabular}
\end{ruledtabular}
\end{table}

\subsection{Behavior of MAE cylinders with different volume fractions at a constant aspect ratio}
\begin{figure}[ht]
	\centering
	\includegraphics[scale=.85]{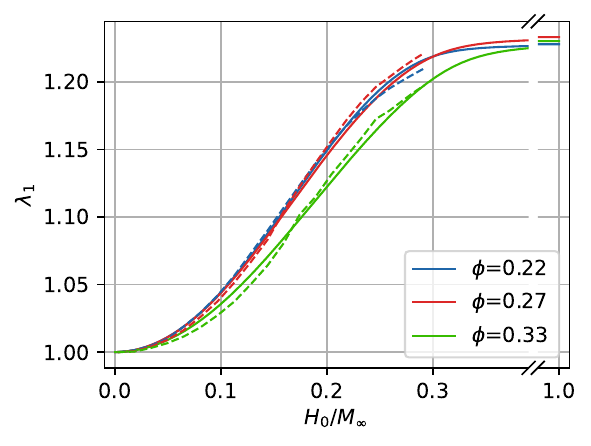}
	\caption{Comparison between the model predictions (solid lines) and the experimental results (dashed lines) for magnetically induced changes in axial stretch ratio ($\lambda_1$) of MAE cylinders with different volume fractions and initial aspect ratio $\gamma_0 =0.6$.}
	\label{p_lz}
\end{figure}

The magnetically induced axial deformation ($\lambda_1$) of cylindrical MAE samples with different volume fractions are shown in Fig. \ref{p_lz}. Here and further, the external magnetic field is normalised by dividing it with the saturation magnetization value of the 
iron particles. The initial aspect ratio of all the cylinders is $\gamma_0 =0.6$. The experimental results presented by dashed lines are the average axial deformation across the upper surface of the cylinder. The model predictions (solid lines) agree very well with the experimental results, which is confirmed by rather low values of the root mean squared error (RMSE): $0.0031$, $0.0049$ and $0.0046$ for the samples with $\phi=0.22, 0.27$ and $0.33$, respectively. The cylinders with $\phi=0.22$ and $0.27$ have a similar deformation behavior, but the one with the highest volume fraction, $\phi=0.33$, exhibits a lower deformation compared to the other two.
The origin of this non monotonic behavior will be explained below when discussing the predicted evolution of the microstructure and its influence on the elastic modulus of the cylinder during actuation. We further observe that the deformation of all samples reach their limit after the external magnetic field exceeds half the value of the saturation magnetization. This corresponds to the field at which the magnetization of particles within the sample comes close to saturation. The other noticeable feature is that the sample with $\phi=0.27$ (red curve) has the highest deformation at saturating fields. The deformation of the highest filled cylinder surpasses the deformation of the cylinder with the lowest volume fraction at saturating fields but still remains lower than for the sample with $\phi=0.27$. This optimal volume fraction has also been found by \citet{diguet2010shape} and \citet{davis1999model}. Although our magnetic field source is limited to $H_0<0.3M_{\infty}$, we know that the magnetic force which causes the field-induced deformation is driven by the magnetization of the sample \cite{romeis2020magnetic}. Therefore, it is natural for the field-induced deformation to follow the trend of the magnetization and reach saturation when the magnetization saturates. This observation is also supported by experiments in high magnetic fields conducted by \citet{diguet2010shape, diguet2009dipolar}.

\begin{figure}[ht]
	\centering
	\includegraphics[scale=.85]{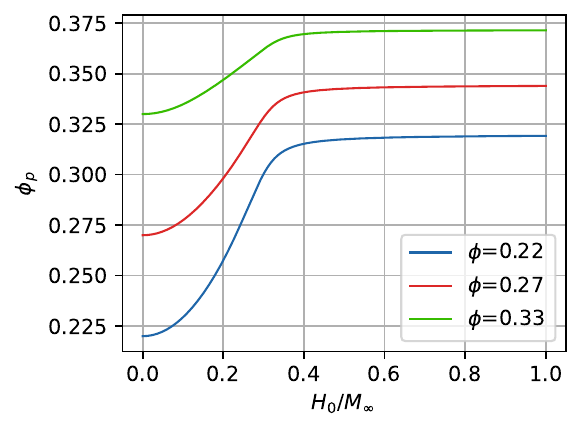}
	\caption{Model predictions for the magnetically induced microstructure evolution (change in particle volume fraction inside the columns ($\phi_p$)) in MAE cylinders with initial aspect ratio $\gamma_0 =0.6$ and various volume fractions.}
	\label{p_p}
\end{figure}

To understand the causes of non monotonic deformational behavior, it is helpful to look into hidden processes taking place inside the MAE sample when it deforms. These processes are the evolution of microstructure, the resulting changes in stiffness and magnetization of the particles, which are only accessible with our model. First, we present the development of columnar structures in MAE cylinders under magnetic loading in Fig. \ref{p_p}. The volume fraction of particles inside the columns $\phi_p$ is obtained as a result of our fit to the experimental deformations shown before.  
It can be seen that the cylinder with the lowest CIP volume fraction has the strongest increase in $\phi_p$ with the increase in magnetic field. In comparison, the two cylinders with higher volume fractions show a weaker microstructure evolution during magnetic loading. This happens because of the stiffening effect of CIP particles on the composite. As the amount of CIP particles within the MAE increases, the effective elastic modulus grows, making it harder for the particles to stretch the elastomeric surroundings and move. This eventually hinders the microstructure evolution for highly filled MAEs. The microstructure evolution stops when the particles magnetization gets close to saturation. This is regardless of the sample's CIP volume fraction.

\begin{figure}[ht]
	\centering
	\includegraphics[scale=.85]{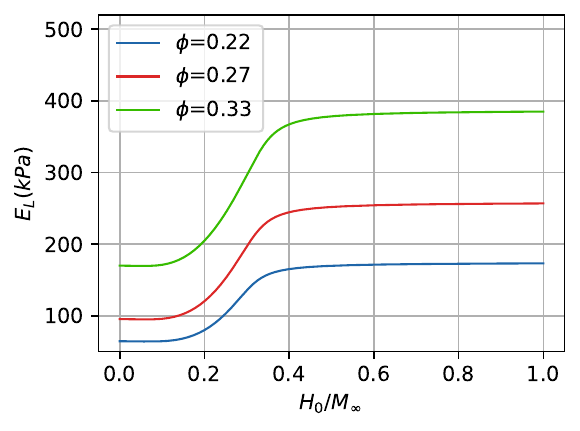}
	\caption{Model predictions for the magnetically induced change of elastic modulus along the field direction ($E_{L}$) (magnetorheological effect) in MAE cylinders with initial aspect ratio $\gamma_0 =0.6$ and various volume fractions.}
	\label{p_e}
\end{figure}

The presented microstructure evolution gives rise to changes in the elastic modulus of the cylinders along the magnetic field direction ($E_L$). This change with respect to the external magnetic field is presented in Fig. \ref{p_e}. 
The lowest filled sample undergoes the smallest increase in $E_L$, even though it had the largest microstructure evolution amongst the samples. Samples with higher volume fractions undergo larger stiffening despite their lower microstructure evolution. The explanation for this lies in the fact that samples with higher CIP filling, although showing a lesser degree of microstructure evolution than their less-filled counterparts, realize a greater amount of CIP within their columnar microstructure. Therefore, $\phi_p$ gets nearer to the maximum packing volume fraction ($\phi_{max}$), at which the effective elastic modulus diverges (see Fig. \ref{EMT}), resulting in a steep increase in the effective elastic modulus of the columns and ultimately intensifying the magnetorheological effect of the more filled samples. Also, we can see that the highest filled cylinder starts with a much higher modulus compared to the less filled cylinders. These two features provide an explanation for the non monotonic behavior observed in elongation results, in which the highest filled cylinder showed lower initial deformation than the other cylinders with lower volume fraction of particles. The elastic forces in the material overpower the magnetic forces, resulting in weaker actuation in the highest filled cylinder. The highest relative increase in the modulus of our cylinders is less than three times their initial modulus and occurs in the case of the lowest filled sample.  

\begin{figure}[ht]
	\centering
	\includegraphics[scale=.85]{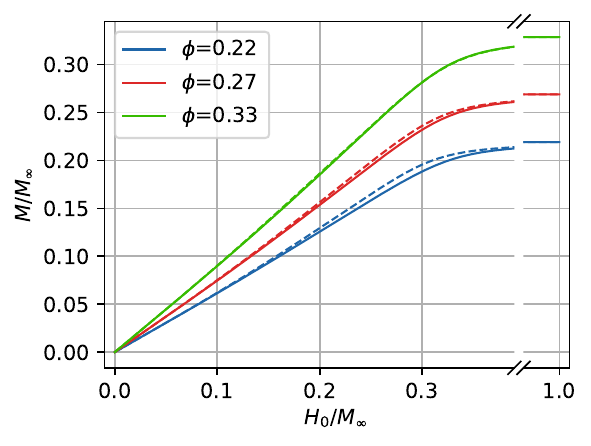}
	\caption{The predicted magnetization of MAE cylinders with initial aspect ratio $\gamma_0 =0.6$ and different volume fractions. The solid lines indicate predictions for the samples with frozen microstructure. The dashed lines represent the magnetization of samples undergoing microstructure evolution.}
	\label{p_m_c}
\end{figure}

Another process that is difficult to measure experimentally during sample deformation but is very important for understanding the behavior of MAEs is the magnetization of the material and its evolution during magnetic field loading. In Fig. \ref{p_m_c} we present two sets of magnetization curves represented with dashed or solid lines.
The dashed lines correspond to the same set of model predictions with evolving microstructure ($k_m=550$). In order to assess the impact of microstructure evolution on the magnetization of the particles, another set of model predictions are presented with a very high micro-mechanical parameter ($k_m\to\infty$) which freezes the microstructure in its initial randomly distributed state during magnetic loading. For a frozen microstructure, shown in solid lines in Fig. \ref{p_m_c}, particle magnetization is not affected by the evolution of the microstructure. It is observed that the lowest filled cylinder that undergoes microstructure evolution (dashed blue line) has a higher slope of magnetization in the transition zone compared to the case with frozen microstructure (solid blue line), before both curves saturate at the same magnetization value. The initial slopes of the sample with frozen microstructure and the sample undergoing microstructure evolution are similar. As the CIP filling of samples increases, this difference decreases and is almost no longer noticeable for the highest filled sample (green lines). This phenomenon is experimentally observed in the case of fully clamped samples \cite{borin2022magneto, bodnaruk2019magnetic}, but the advantage here is that we observe it in deforming samples.

\begin{figure}[ht]
	\centering
	\includegraphics[scale=.85]{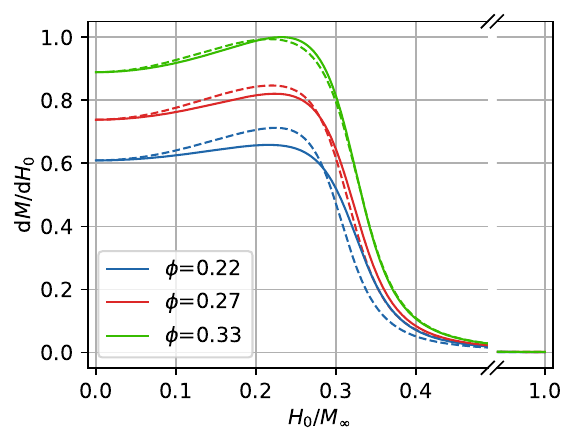}
	\caption{The differential susceptibility of MAE cylinders as a derivative of the magnetization with respect to the external magnetic field. The samples with the initial aspect ratio $\gamma_0 =0.6$ contain different volume fractions of particles. The solid lines belong to the model evaluation for the samples with frozen microstructure. The dashed lines represent the differential susceptibility of samples undergoing microstructure evolution.}
	\label{dp_m_c}
\end{figure}

By taking the derivative of the magnetization $M$ with respect to the external magnetic field $H_0$, we obtain the differential susceptibility of the MAE cylinders, as shown in Fig. \ref{dp_m_c}. 
As in the previous figure, the solid lines represent the samples with frozen microstructure. The dashed lines indicate the predicted values based on our fit to the experimental deformation results. The observed uptake in all the solid lines in Fig. \ref{dp_m_c} is due to the sample elongation which decreases the demagnetizing factor (Eq. \ref{demag}) and hence increases the magnetization. The difference between the dashed and solid lines for each volume fraction essentially reflects the impact of microstructure evolution on the differential susceptibility of the samples. This difference is greatest for the sample with the lowest filler content (blue lines in Fig. \ref{dp_m_c}) and decreases as filler content increases. This directly correlates with the extent of microstructure evolution in the samples shown in Fig. \ref{p_p}; the more the sample is filled with particles, the more the microstructure evolution is hindered. From this we can conclude that a pronounced microstructure evolution in an MAE cylinder leads to a strong increase in magnetic susceptibility within the transition zone of the magnetization curves.

\subsection{Behavior of cylinders with different aspect ratios at a constant CIP volume fraction}
\begin{figure}[ht]
	\centering
	\includegraphics[scale=.85]{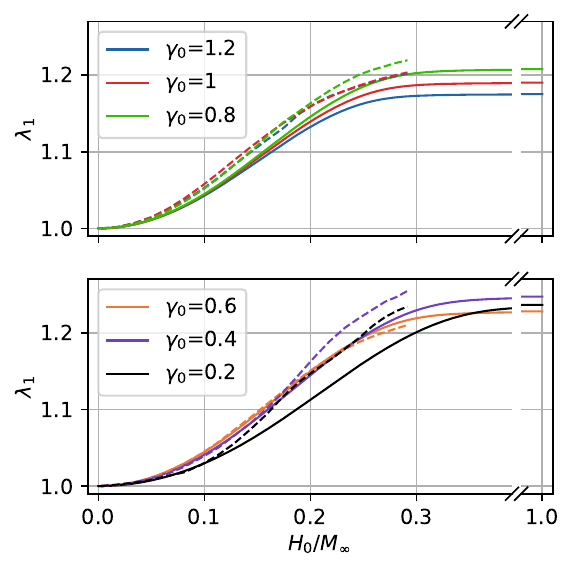}
	\caption{Comparison between the model predictions (solid lines) and experimental results (dashed lines) for magnetically induced changes in the axial stretch ratio $\lambda_1$ of MAE cylinders with different aspect ratios and with the volume fraction $\phi=0.22$.}
	\label{g_lz}
\end{figure}

Above, we focused on the behavior of MAE cylinders with different volume fractions at one aspect ratio. Further, we present the results of our study for samples with different aspect ratios while keeping the volume fraction of the samples at $\phi=0.22$. In Fig. \ref{g_lz}, the magnetically induced axial deformation $\lambda_1$ of MAE cylinders with respect to the external magnetic field is presented. Dashed lines correspond to the experimental results, while the model predictions are plotted with solid lines. The RMSE values indicate good agreement between theory and experiment: $0.0195$, $0.0153$, $0.0135$, $0.0031$, $0.0167$ and $0.0258$ for the samples with $\gamma_0=1.2, 1, 0.8, 0.6, 0.4$ and $0.2$, respectively. To avoid cluttering the figure, we divide the results into two sets according to the aspect ratios shown in the top and bottom subplots.
The experimental results and model predictions both show that the samples have a similar initial deformation slope with respect to $H_0$, except for the cylinder with $\gamma_0=0.2$, which has a smaller initial slope compared to the others. Regarding the deformation at high magnetic fields, we see that the trend in both experiment and our model point to an increase in the axial stretch ratio when $\gamma_0$ is decreased. Here again, the sample with the initial aspect ratio $\gamma_0=0.2$ shows a deviation from the trend, both in the experimental results and in the model predictions, as it has lower deformation in moderate fields compared to the other samples.

Fig. \ref{g_lz} also indicates that by increasing the initial aspect ratio ($\gamma_0$) of the cylinder, the deformation limit is reached at lower magnetic fields compared to samples with a smaller $\gamma_0$. This deformation limit is reached for the cylinder with the smallest aspect ratio ($\gamma_0=0.2$) when the external magnetic field reaches about half of the saturation magnetization value. For the sample with the largest aspect ratio ($\gamma_0=1.2$), this limit occurs when the external magnetic field reaches about one-third of the saturation magnetization. The deformation of MAE cylinders in saturating magnetic fields is also larger for samples with smaller initial aspects ratios except in the case of the sample with $\gamma_0=0.2$.

\begin{figure}[ht]
	\centering
	\includegraphics[scale=.85]{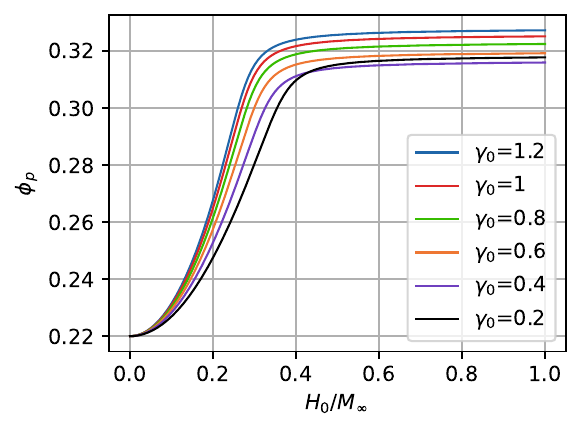}
	\caption{Model predictions for the magnetically induced microstructure evolution (change in particle volume fraction inside the columns ($\phi_p$)) in MAE cylinders with the initial aspect ratio $\gamma_0 =0.6$ and various volume fractions.}
	\label{g_p}
\end{figure}

The development of the particle volume fraction within the columns ($\phi_p$) while the samples undergo magnetically induced deformation is presented in Fig. \ref{g_p}.
Here, we see a steeper initial slope of microstructure evolution for samples with a higher initial aspect ratio ($\gamma_0$). It is known that more elongated shapes have a smaller demagnetizing field and thus the magnetization field is increased. Therefore, the dipolar forces between the particles are higher in samples with larger $\gamma_0$, which results in denser columns and steeper microstructure evolution, as seen in this figure. Also, the value of $\phi_p$ at saturating magnetic fields is larger for samples with a larger $\gamma_0$, again except for the sample with $\gamma_0=0.2$.

\begin{figure}[ht]
	\centering
	\includegraphics[scale=.85]{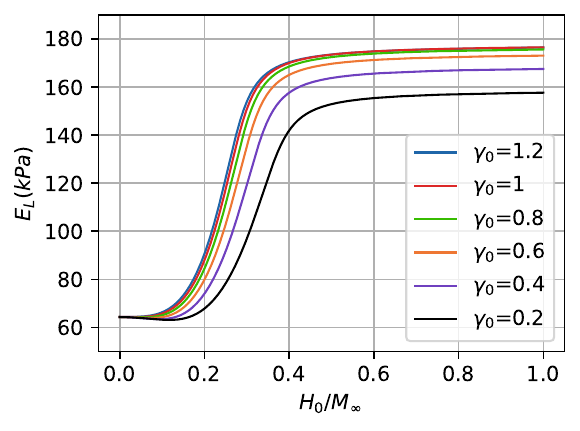}
	\caption{The effect of sample aspect ratio ($\gamma$) on the magnetically induced change of elastic modulus ($E_L$) along the field direction (magnetorheological effect) in MAE cylinders with volume fraction $\phi=0.22$, based on our presented model.}
	\label{g_e}
\end{figure}

The changes in the elastic modulus of the MAE cylinders resulting from the evolution of their microstructure are presented in Fig. \ref{g_e} with respect to the external magnetic field. 
The first noticeable feature of this figure is that for the cylinders with small initial aspect ratios, the elastic modulus $E_L$ in the direction of magnetic field slightly decreases with increasing external magnetic field before increasing again. This negative rheological effect, which was also observed in our previous studies \cite{chougale2021transverse}, is most clearly visible for the sample with $\gamma_0=0.2$ and could be a reason for the anomalous behaviour of this sample seen discussed above. The value of $E_L$ at saturating magnetic field shows an increasing trend for higher initial aspect ratios until $\gamma_0=0.8$, after which an increase in $\gamma_0$ does not seem to noticeably affect the behavior of MAE cylinders. The maximum magnetorheological effect observed here is a threefold increase in elastic modulus for the cylinder with the largest initial aspect ratio. 

Despite the experimental limitations that keep us from measuring the modulus of freely deforming cylinders in this study, the magnitude of the magnetorheological effect predicted by our model in Figs. \ref{p_e} and \ref{g_e} is in agreement with studies that use other methods of measuring the modulus of MAEs in a magnetic field. For example, \citet{vatandoost2020dynamic} investigated the dynamic properties of isotropic and anisotropic cylindrical MAEs in compression mode with a large pre-strain. They used cylindrical oblate samples with an aspect ratio of $4/9$ and found a maximum magnetorheological effect for their isotropic samples around 340\% for small strains and 188\% for large strains. \citet{borin2022magneto} used a torsional oscillation method on rod shaped MAEs and found a magnetorheological effect of around 460\% for samples made with carbonyl iron particles. The concentration of particles in their samples was $\sim40$ vol$\%$, and the initial effective Young's modulus $\sim$ 120 - 140 kPa (G $\sim$ 40-46 kPa), which are similar to the material parameters in the present paper.

\section{Conclusions}

This collaborative study employs a novel model that considers the effects of microstructure evolution on the mechanical behavior of MAE cylinders and combines its predictions with experimental measurements. This give us a unique opportunity to understand in depth how microstructural evolution affects the mechanical and magnetic behavior of MAEs. We use the dipolar mean-field approach for taking the magnetic interactions at micro- and macroscale into account. A penalty term is added to the transversely isotropic Neo-Hookean elastic model to account for the energy spent on particle restructuring within the elastomeric matrix. Eight cylindrical MAE samples, with various aspect ratios and CIP volume fractions, are fabricated and tested in a uniform magnetic field. By fitting the model to experimental data, our study allows us to simultaneously access the changes in three main MAE properties: deformation, stiffness and magnetization. 
This provides a better understanding of the magnetorheological phenomenon in MAEs. Multiple conclusions can be drawn from the results as follows:

\begin{enumerate}
\item The axial deformation predictions of our model agree quantitatively with experimental data for MAE cylinders with different volume fractions and the same aspect ratio. Also for samples that differ in their aspect ratio but have the same volume fraction, the model is in a very good qualitative agreement with the experiment and successfully predicts the major trends. 
\item Both the model and the experiment show that there is a limit to how much an increase in the volume fraction can improve the behaviour of the material in case of deformation. The optimal volume fraction is found to be around $\phi=0.27$. From the insights our model provides into microstructure and stiffness changes, we understand that this limitation comes from the overpowering of magnetic forces by amplified elastic forces resulted from enhanced microstructure evolution and stiffness of the material in highly filled samples.
\item Comparing samples with different aspect ratios and constant CIP volume fraction, we find that the maximum deformation is larger for samples with a lower aspect ratio. This is explained by looking at the microstructure evolution which is stronger at higher aspect ratios, leading to higher modulus and consequent lower deformation for more elongated samples.
\item This work suggests that the magnetorheological effect in soft (shear modulus of the empty matrix is somewhat below 10 kPa and the resulting effective shear modulus of isotropic MAEs is $\sim30 - 50$ kPa) isotropic MAE materials can not exceed one order of magnitude in the case of unconstrained (free at one end) samples in a uniform magnetic field. The strongest magnetorheological effect predicted by the unified mean-field model here is less than threefold and occurs for the cylinder with the lowest CIP volume fraction. Any higher value would result in diminished deformation, and that is not observed experimentally. 
\end{enumerate}

From a fundamental point of view, it would be interesting to verify experimentally the hypothesis (4) by measuring the apparent elastic and shear moduli of deformed macroscopic cylinders in external magnetic fields. Additionally, a theoretical study on the shearing of MAE discs that accounts for microstructure evolution would greatly enhance our understanding of the behavior of these materials.

\bibliography{main.bib}

\end{document}